\documentclass[a4paper]{article}

\usepackage{INTERSPEECH2022}
\usepackage{cite}
\usepackage{booktabs}
\usepackage[hidelinks]{hyperref}
\usepackage{xurl}
\usepackage{makecell}
\usepackage{multirow}
\usepackage{xcolor}
\usepackage{colortbl}

\definecolor{highlight}{HTML}{ade6dc}

\newcommand{\lang}[1]{\texttt{#1}}
\newcommand{\loc}[2]{\makecell{\texttt{#1}\\\texttt{#2}}}
\newcommand{\best}[1]{\begingroup\setlength{\fboxsep}{0.8pt}\colorbox{highlight}{\textbf{#1}}\endgroup}

\newcommand{\urlfootnote}[1]{%
  \footnote{%
    \begingroup
    \scriptsize
    \raggedright
    \setlength{\emergencystretch}{2em}%
    \hangindent=1.2em
    \hangafter=1
    \url{#1}%
    \endgroup
  }%
}

\title{HINTT Submission to the 2nd MLC-SLM Challenge:\\Comparing Cascaded and Unified Approaches to Diarization and ASR}
\name{Takanori Ashihara, Kohei Matsuura, Masato Mimura}
\address{NTT, Inc., Japan}
\email{takanori.ashihara@ntt.com}

\begin{document}

\maketitle

\begin{abstract}
This paper presents the HINTT system submitted to the 2nd Challenge and Workshop on Multilingual Conversational Speech Language Model (MLC-SLM).
We address multilingual speaker-attributed ASR, where systems must determine who spoke when and what was spoken.
We investigate two modeling strategies for this problem: a cascaded pipeline that combines speaker diarization with speech-LLM-based ASR, and a unified speech LLM that directly generates speaker labels, timestamps, and transcriptions.
Our final submission is based on the cascaded pipeline, consisting of a fine-tuned DiariZen diarization model, a fine-tuned Qwen3-ASR model, and LLM-based generative error correction.
For comparison, we also fine-tune VibeVoice-ASR as a unified model using the same official training data.
All task-specific fine-tuning and model selection are performed using only the official MLC-SLM data, without external data or pseudo-labels.
Experimental results demonstrate that the cascaded system remains more reliable under the MLC-SLM Task 1 conditions, while unified speech LLMs offer a promising direction for future speaker-attributed ASR.
\end{abstract}

\section{Introduction}
Understanding ``who spoke what and when'' is a fundamental requirement for conversational speech processing.
This problem goes beyond conventional automatic speech recognition (ASR), because a system must jointly determine speaker turns, assign each utterance to the correct speaker, and transcribe the corresponding speech content.
Such speaker-attributed transcription is essential for meeting transcription, conversation analysis, spoken document retrieval, and downstream spoken language understanding~\cite{saasr_compara,survey_multi_asr}.
Accordingly, the joint problem of speaker diarization and ASR has been extensively studied in both research benchmarks and community challenges, including the CHiME challenges~\cite{chime6,chime7,chime8}.

The 2nd Challenge and Workshop on Multilingual Conversational Speech Language Model (MLC-SLM) further highlights this problem in the context of multilingual speech large language models (LLMs).
In Task 1, systems are required to perform multilingual conversational speech diarization and recognition without oracle utterance boundaries or speaker labels.
This setting is challenging because the system must handle not only multilingual and accented speech, but also speaker attribution in natural two-speaker conversations.
The challenge permits both pipeline-based and end-to-end systems, and evaluates submissions using time-constrained minimum permutation word/character error rate (tcpWER/tcpCER), with the final ranking based on the averaged value across languages.

A conventional and widely used approach to this problem is a cascaded system, in which speaker diarization and ASR are handled by separate modules.
Such modular systems are attractive because each component can be independently optimized, replaced, and analyzed.
Accordingly, they have been widely adopted in previous challenge submissions, including those to the CHiME series~\cite{chime6,chime7,chime8,chime8_ntt}.
A similar trend was observed in the first MLC-SLM challenge~\cite{mlcslm2025}, where strong published systems combined dedicated diarization models with ASR~\cite{1st_ranked_mlcslm2025,unified_mlcslm2025}.
These findings indicate that cascaded systems remain a strong and practical baseline for multilingual speaker-attributed ASR.

In contrast, recent unified speech-LLM-based models attempt to jointly model speaker diarization and ASR for multi-speaker long-form speech~\cite{speakerlm,jedis_llm,tagspeech,speaker_reasoner,vibevoice_asr}.
For example, VibeVoice-ASR~\cite{vibevoice_asr} formulates long-form speech recognition as a unified speech-to-text generation task that outputs speaker labels, timestamps, and transcriptions.
This line of work is appealing because it can reduce hand-crafted interfaces between modules and potentially exploit long-range conversational context.
However, despite recent progress in unified models, it remains underexplored whether they can match or outperform carefully tuned cascaded systems in the MLC-SLM Task 1 setting.

In this paper, we present the HINTT submission from NTT Human Informatics Laboratories to the 2nd MLC-SLM Challenge and compare cascaded and unified approaches for multilingual conversational speech recognition with speaker diarization.
Our submitted system follows a cascaded design: a fine-tuned speaker diarization model first estimates speaker-labeled segments, and a fine-tuned speech-LLM-based ASR model then transcribes each segment.
We further apply generative error correction (GEC)~\cite{gec_llm,hyporadise,nbest_t5,gensec,rllm_cf} using an LLM to refine the ASR hypotheses.
In addition, we compare this cascaded system with a fine-tuned VibeVoice-ASR model as a unified counterpart.
Experimental results show that the cascaded system provides a more reliable submission under the challenge conditions, while the unified model remains promising, despite being fine-tuned only on about 60-s chunks due to computational constraints.

\begin{figure}[t]
  \centering
  \includegraphics[width=1.0\linewidth]{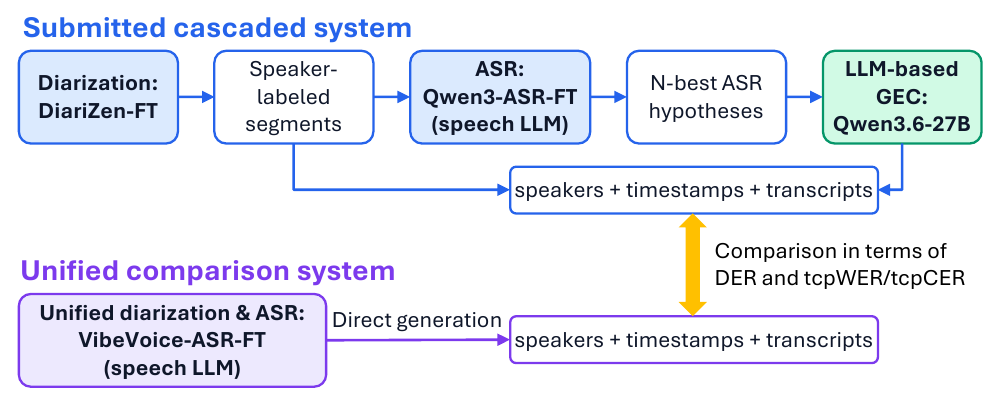}
  \caption{Overview of the cascaded and unified systems.}
  \label{fig:system}
\end{figure}

\section{System overview}
\label{sec:system}
Figure~\ref{fig:system} shows an overview of the systems investigated in this work.
We compare two different approaches to multilingual conversational speech diarization and recognition.
The first is a cascaded approach, where speaker diarization and ASR are performed by separate models.
The second is a unified approach, where a single speech-LLM-based model directly predicts speaker labels, timestamps, and transcriptions.

For training and validation, we used only the official training and development splits provided by the 2nd MLC-SLM Challenge.
The training dataset comprises approximately 2\,100 hours of multilingual conversational speech covering diverse languages and accents.
We did not use any external speech/text data or pseudo-labels for fine-tuning.
All experiments, including task-specific fine-tuning, decoding, and LLM-based GEC, were conducted under a bounded compute setting using at most four NVIDIA RTX 6000 Ada GPUs, each with 48 GB of VRAM.

To alleviate a dataset mismatch, we applied silence masking to both the training and development splits.
Specifically, all audio regions outside the annotated speech intervals were replaced with silence.
This preprocessing was motivated by the observation reported in the first MLC-SLM Challenge that the evaluation recordings appeared to be silence-masked outside the labeled speech regions~\cite{but_mlcslm2025}.
We also observed a similar tendency in the current challenge setting.
In addition, our manual inspection by listening to the original training and development recordings showed that some regions outside the annotated intervals contained audible speech.
This mismatch is less problematic for utterance-level ASR models, which are trained and evaluated on already segmented speech.
However, it can severely affect diarization models and unified diarization--ASR models, as these models may learn to treat actual speech outside the annotations as non-speech.
We therefore used the silence-masked training and development splits to reduce this mismatch.
Accordingly, all development-set results reported below were computed on this modified development split.

\subsection{Submitted cascaded system}
\label{subsec:cascade}
\textbf{Approach:}
Our submitted system follows a cascaded diarization--ASR design, as illustrated in the upper part of Fig.~\ref{fig:system}.
First, a fine-tuned speaker diarization model estimates speaker-labeled speech segments from the input conversation.
We use DiariZen~\cite{diarizen} as the diarization backend, which has shown strong performance in the first MLC-SLM Challenge~\cite{but_mlcslm2025}.
During inference, the diarization output is converted into a list of speaker-attributed segments, which are then used to divide the input audio.

Each diarized segment is subsequently recognized by a fine-tuned speech-LLM-based ASR model.
In this work, we use Qwen3-ASR~\cite{qwen3_asr} as the ASR backend and fine-tune it using language-only prompts, without accent labels.
Since the ASR model operates on individual diarized segments, the cascade explicitly decomposes the overall task into ``who spoke when'' and ``what was spoken.''
This modular design allows the diarization and ASR components to be optimized independently.
It also enables the use of a strong segment-level ASR model~\cite{qwen3_asr,whisper} without training it on full-session audio, thereby substantially reducing the training cost.

For Qwen3-ASR inference, we use the same language prompts as in training.
Each diarized speech segment is decoded without sampling using a beam size of 5, and the 5-best hypotheses are retained together with their sequence scores for the subsequent GEC stage.
During decoding, we observe that the autoregressive decoder occasionally produces generation loops with repeated outputs.
To handle such cases, we incorporate a repetition-detection and recovery procedure inspired by the VibeVoice auto-recovery script.\urlfootnote{https://github.com/microsoft/VibeVoice/blob/main/vllm_plugin/tests/test_api_auto_recover.py}
Specifically, we detect repetition loops in the top hypothesis using two criteria.
First, we check whether a substring of at least 10 characters is repeated consecutively at least 10 times within the last 400 characters of the output.
Second, we check whether the same word n-gram is repeated consecutively at least ten times anywhere in the output, considering n-grams from unigrams to five-grams.
A segment is re-decoded only when a repetition loop is detected, using a sampling-based temperature-escalation schedule with temperatures of 0.2, 0.3, and 0.4 and top-$p=0.95$.
This recovery path was triggered for only a few utterances in the evaluation set.

Finally, we apply GEC to the N-best ASR hypotheses using an LLM~\cite{gec_llm,hyporadise,nbest_t5,gensec,rllm_cf}.
The GEC module is prompted to choose the most plausible hypothesis or correct obvious recognition errors when necessary.
Since GEC is used only as text post-processing, the final output combines the diarization-derived speaker labels and timestamps with the corrected transcriptions.

\noindent\textbf{Training and inference configuration:}
For DiariZen, we fine-tuned the public checkpoint\urlfootnote{https://huggingface.co/BUT-FIT/diarizen-wavlm-large-s80-md} following the official \texttt{diar\_ssl} recipe.
We trained the model for up to 30 epochs, with learning rates of $1\times10^{-5}$ for the WavLM frontend and $1\times10^{-4}$ for the Conformer backend, a weight decay of 0.01, an effective batch size of 16, and an early-stopping patience of 5.
Since the chunk size can affect diarization performance~\cite{eend_chunk_effect1,eend_chunk_effect2,eend_chunk_effect3}, we used 64-s chunks with a 48-s shift for training.

For DiariZen inference, we used a checkpoint averaged over the three checkpoints with the lowest validation loss and applied the DiariZen inference pipeline with the same 64-s chunk duration.
In the official DiariZen implementation, the 64-s chunk duration also affects which speaker embeddings are retained for clustering, potentially leading to more reliable speaker clusters~\cite{eend_chunk_filter}.
For speaker clustering, we adopted agglomerative hierarchical clustering with ResNet34-based speaker embeddings.\urlfootnote{https://huggingface.co/pyannote/wespeaker-voxceleb-resnet34-LM}
Since each input recording always contained two speakers, we fixed the number of speakers to two and used a clustering threshold of 0.7 and a minimum cluster size of 16.
We denote this model as DiariZen-FT and compare it with the original DiariZen using the default inference settings.
We also evaluate a variant using a SimAMResNet100-based speaker embedding model trained on the multilingual VoxBlink2 corpus~\cite{voxblink2} and provided by WeSpeaker.\urlfootnote{https://github.com/wenet-e2e/wespeaker/blob/master/docs/pretrained.md}
In addition, we compare it with Baseline$^{\mathrm{1st}}$, which uses the baseline recipe from the first challenge~\cite{mlcslm2025}\urlfootnote{https://github.com/mubingshen/MLC-SLM-Baseline} but was trained on the second-challenge dataset.

For Qwen3-ASR, we adapted the official 1.7B-parameter checkpoint\urlfootnote{https://huggingface.co/Qwen/Qwen3-ASR-1.7B} using the official \texttt{finetuning} recipe.
Optimization was performed for up to 5 epochs with a learning rate of $5\times10^{-5}$ and an effective batch size of 128, while the remaining hyperparameters were kept unchanged from the official recipe (hereafter, Qwen3-ASR-FT).
We further applied SpecAugment~\cite{specaugment} during fine-tuning, aiming to make the ASR model less sensitive to local acoustic dropouts caused by diarization-stage segmentation errors.
Specifically, SpecAugment was applied to the log-Mel input features produced by the Qwen3-ASR processor before they were fed into the Qwen3-ASR audio encoder.
We used two time masks with a maximum width of 40 frames and two frequency masks with a maximum width of 12 mel bins, setting the masked regions to zero.

\begin{figure}[t]
  \centering
  \includegraphics[width=1.0\linewidth]{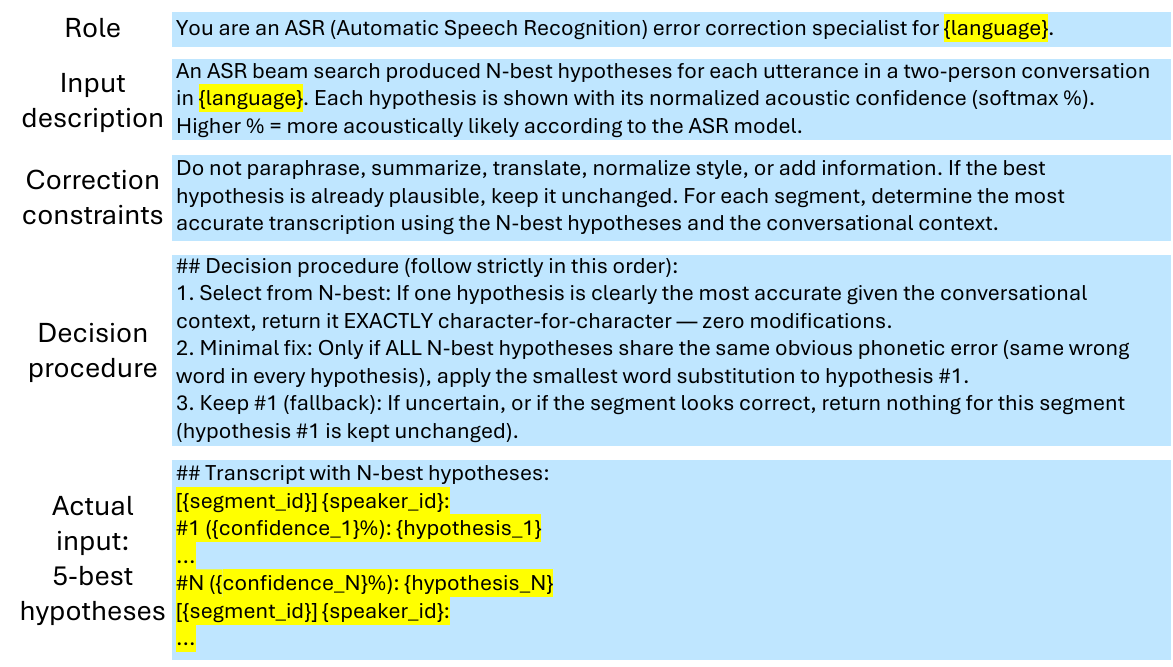}
  \caption{Prompt template for LLM-based GEC. Highlighted fields are replaced with the target language and N-best hypotheses with softmax-normalized beam-search sequence scores.}
  \label{fig:gec_prompt}
\end{figure}

For the GEC stage, we used the original Qwen3.6-27B checkpoint\urlfootnote{https://huggingface.co/Qwen/Qwen3.6-27B} without fine-tuning.
As shown in Fig.~\ref{fig:gec_prompt}, the LLM was given a session-level transcript containing the 5-best hypotheses from Qwen3-ASR-FT, together with their softmax-normalized beam-search sequence scores.
The model was instructed to perform selection-first correction: it selected a non-top hypothesis only when supported by the conversational context, applied minimal corrections only for shared phonetic errors, and kept the top hypothesis unchanged when uncertain.

\subsection{Unified comparison approach}
\label{subsec:unified}
\textbf{Approach:}
In addition to the cascaded system, we investigate a unified speech-LLM-based approach using VibeVoice-ASR~\cite{vibevoice_asr}, as shown in the bottom part of Fig.~\ref{fig:system}.
Unlike the cascaded system, VibeVoice-ASR directly generates speaker-attributed transcriptions with timestamps from long-form audio.
This formulation is attractive because it removes the hand-crafted interface between diarization and ASR, while allowing the model to exploit acoustic, lexical, and conversational context in a single generation process.

Ideally, such a model would be fine-tuned on complete long-form sessions.
However, full-session fine-tuning is computationally infeasible under our hardware budget, and we therefore fine-tune it in a chunk-wise manner.
Accordingly, for long-form inference with the fine-tuned model, we initially adopted a simple strategy that decodes overlapping chunks independently and then merges the hypotheses.
However, for VibeVoice-ASR, which outputs speaker-attributed JSON sequences, we found this strategy unreliable: speaker labels were often assigned inconsistently across decoded chunks, making it difficult to maintain speaker identity over the full session.

Instead, we use a prefix-conditioned continuation strategy inspired by previous-output conditioning in long-form ASR systems such as Whisper~\cite{whisper}.
Unlike text-only prompting, our prefix consists of VibeVoice-ASR JSON outputs from previous chunks, including timestamps, speaker labels, and transcriptions.
By conditioning the next chunk on this structured history, the model can continue generation from the established speaker and timestamp context, rather than assigning speaker labels independently for each chunk.

Repetition loops occur substantially more frequently with VibeVoice-ASR than with Qwen3-ASR.
Therefore, as in Sec.~\ref{subsec:cascade}, we also apply repetition-loop detection during generation.
When a repetition loop is detected, decoding is retried in two stages.
First, we change only the random seed while keeping the decoding procedure deterministic, i.e., without sampling.
This is effective because VibeVoice-ASR applies a random perturbation, similar to dithering, before the acoustic representations are passed to the LLM; therefore, changing the seed can alter the generated sequence even without sampling.
In practice, this seed-based retry recovered several chunks from repetition loops.
If the loop still persists, we then enable temperature-based sampling as a fallback.

\noindent\textbf{Training and inference configuration:}
We fully fine-tuned the official VibeVoice-ASR checkpoint\urlfootnote{https://huggingface.co/microsoft/VibeVoice-ASR}.
During training, each long-form recording was split into approximately 60-s chunks based on oracle utterance boundaries.
We trained the model for one epoch with a learning rate of $5\times10^{-5}$, a warmup ratio of 0.03, a weight decay of 0.01, and an effective batch size of 128 (hereafter, VibeVoice-ASR-FT).

At inference time, each recording was decoded using 60-s chunks with prefix-conditioned continuation.
For each chunk, previously committed VibeVoice-ASR JSON segments were provided as an assistant prefix, and the model generated the continuation for the current chunk.
Because the final generated segment in each chunk could be truncated at the chunk boundary, we discarded the last segment before committing the generated segments.
The next chunk was then determined from the end time of the penultimate generated segment: its start time was set to 40~s before that endpoint.
This resulted in an adaptive sliding-window procedure with approximately 40-s overlap and 20-s shift, while aligning the chunk transition to the model's predicted segment boundaries.

\begin{table*}[t]
\caption{Diarization performance on the silence-masked development set, evaluated using DER (\%) with a 0-s collar.}
\label{tab:res_der}
\centering
\setlength{\tabcolsep}{1.5pt}
\scriptsize
\resizebox{\textwidth}{!}{%
\begin{tabular}{@{}ll*{22}{c}@{}}
\toprule
ID
& System
& \loc{en-}{US}
& \loc{en-}{AU}
& \loc{en-}{GB}
& \loc{en-}{PH}
& \loc{en-}{IN}
& \lang{fr}
& \loc{fr-}{CA}
& \lang{de}
& \lang{it}
& \lang{pt}
& \loc{pt-}{BR}
& \lang{ru}
& \lang{es}
& \loc{es-}{MX}
& \lang{tl}
& \lang{tr}
& \lang{ur}
& \lang{vi}
& \lang{ja}
& \lang{ko}
& \lang{th}
& Avg. \\
\midrule
\texttt{D1}
& Baseline$^{\mathrm{1st}}$
& 11.32 & 11.92 & 15.44 & 13.57 & 8.17
& 11.70 & 18.95 & 10.96 & 9.89 & 10.92
& 9.58 & 9.27 & 9.04 & 7.09 & 13.87
& 16.08 & 9.64 & 7.22
& 19.84 & 14.88 & 10.17
& 11.88 \\
\midrule
\texttt{D2}
& DiariZen
& 13.91 & 18.41 & 13.19 & 15.27 & 8.32
& 20.93 & 16.66 & 17.60 & 10.67 & 18.57
& 8.89 & 13.20 & 15.06 & 13.54 & 16.25
& 33.40 & 20.63 & 14.55
& 27.15 & 23.82 & 17.86
& 17.04 \\
\texttt{D3}
& DiariZen-FT
& 2.29 & \best{2.02} & \best{3.75} & 6.33 & \best{0.79}
& \best{2.29} & \best{8.86} & 1.70 & 2.24 & 4.11
& 0.77 & 2.25 & 3.30 & \best{1.05} & 1.60
& 9.47 & 2.31 & \best{2.05}
& 3.32 & 5.15 & 2.63
& 3.25 \\
\texttt{D4}
& \hspace{1em}+ SimAMResNet100
& 2.30 & \best{2.02} & \best{3.75} & 6.44 & 0.98
& \best{2.29} & 8.95 & 1.68 & 2.24 & \best{2.77}
& 0.77 & \best{2.25} & \best{3.28} & \best{1.05} & 1.59
& \best{4.08} & 2.31 & 2.18
& \best{2.90} & \best{5.11} & \best{2.60}
& \best{2.93} \\
\midrule
\texttt{D5}
& VibeVoice-ASR-FT
& \best{2.20} & 2.54 & 5.39 & \best{6.31} & 10.61
& \best{2.29} & 10.41 & \best{1.58} & \best{2.22} & 7.64
& \best{0.41} & 8.16 & 4.13 & 3.76 & \best{1.48}
& 8.64 & \best{1.81} & 7.67
& 8.83 & 7.61 & 3.82
& 5.12 \\
\bottomrule
\end{tabular}%
}
\end{table*}

\begin{table*}[t]
\caption{Overall performance on the silence-masked development set, evaluated with a 5-s collar. Baseline$^\mathrm{2nd}$ (i.e., LoRA-adapted VibeVoice-ASR) results are taken from the official scores. \texttt{S6} is shown only as an ablation and is not used in the submitted system.}
\label{tab:res_tcpwer}
\centering
\setlength{\tabcolsep}{1.5pt}
\scriptsize
\resizebox{\textwidth}{!}{%
\begin{tabular}{@{}ll*{22}{c}@{}}
\toprule
&
& \multicolumn{18}{c}{tcpWER (\%)}
& \multicolumn{3}{c}{tcpCER (\%)}
& \\
\cmidrule(lr){3-20}
\cmidrule(lr){21-23}
ID
& System
& \loc{en-}{US}
& \loc{en-}{AU}
& \loc{en-}{GB}
& \loc{en-}{PH}
& \loc{en-}{IN}
& \lang{fr}
& \loc{fr-}{CA}
& \lang{de}
& \lang{it}
& \lang{pt}
& \loc{pt-}{BR}
& \lang{ru}
& \lang{es}
& \loc{es-}{MX}
& \lang{tl}
& \lang{tr}
& \lang{ur}
& \lang{vi}
& \lang{ja}
& \lang{ko}
& \lang{th}
& Avg. \\
\midrule
\texttt{S1}
& Baseline$^\mathrm{2nd}$
& 77.39 & 81.50 & 67.60 & 63.36 & 72.12
& 83.39 & 78.56 & 84.23 & 78.16 & 75.64
& 73.02 & 83.84 & 82.51 & 78.81 & 81.09
& 92.97 & 89.63 & 71.81
& 81.46 & 81.33 & 83.67
& 79.15 \\
\midrule
\texttt{S2}
& \texttt{D4}$\Rightarrow$Qwen3-ASR-FT
& 13.25 & 8.45 & 12.47 & 13.80 & 12.90
& 16.62 & 31.46 & 18.45 & 14.43 & 23.15
& 10.75 & 15.03 & 12.53 & 7.98 & 24.73
& 23.61 & 17.18 & 13.78
& 16.84 & 16.01 & 12.52
& 16.00 \\
\texttt{S3}
& \hspace{1em}+ SpecAug
& 13.23 & 8.34 & 12.37 & 13.63 & \best{12.78}
& 16.16 & 30.38 & 17.97 & 13.82 & 22.47
& 10.32 & 14.56 & 12.53 & 7.61 & 23.66
& 22.98 & 16.67 & 12.68
& 16.43 & 15.56 & 11.50
& 15.51 \\
\texttt{S4}
& \hspace{1em}++ GEC
& \best{13.17} & \best{8.28} & \best{12.26} & 13.52 & 12.85
& \best{16.05} & \best{30.25} & \best{17.83} & \best{13.50} & \best{22.20}
& \best{9.97} & \best{13.99} & 12.34 & \best{7.38} & \best{23.08}
& \best{22.30} & 16.44 & \best{12.40}
& \best{16.03} & \best{15.38} & 11.56
& \best{15.28} \\
\midrule
\texttt{S5}
& VibeVoice-ASR-FT
& 13.24 & 8.87 & 16.70 & \best{13.37} & 24.48
& 16.66 & 31.81 & \best{17.83} & 15.15 & 28.98
& 10.11 & 21.73 & \best{11.53} & 12.06 & 24.65
& 29.17 & \best{15.98} & 19.88
& 25.47 & 19.57 & \best{11.23}
& 18.50 \\
\midrule
\texttt{S6}
& \texttt{D4}$\Rightarrow$Whisper-FT
& 13.62 & 8.73 & 13.37 & 13.86 & 11.79
& 17.80 & 30.98 & 19.42 & 14.19 & 21.16
& 10.22 & 15.07 & 12.81 & 6.48 & 22.67
& 20.94 & 22.86 & 20.45
& 19.59 & 15.49 & 15.08
& 16.50 \\
\bottomrule
\end{tabular}%
}
\end{table*}

\begin{table}[t]
\caption{DER breakdown. Miss, FA, and CNF denote missed speech, false alarm speech, and speaker confusion, respectively.}
\label{tab:diar_anal}
\centering
\scriptsize
\resizebox{\linewidth}{!}{%
\begin{tabular}{@{}llcccc@{}}
\toprule
ID & Model & Miss (\%) & FA (\%) & CNF (\%) & DER (\%) \\
\midrule
\texttt{D1} & Baseline$^{\mathrm{1st}}$ & 2.39 & 5.64 & 3.85 & 11.88 \\
\texttt{D2} & DiariZen & 7.73 & 3.00 & 6.30 & 17.04 \\
\texttt{D3} & DiariZen-FT & \best{0.06} & \best{0.60} & 2.59 & 3.25 \\
\texttt{D4} & \hspace{1em}+ SimAMResNet100 & \best{0.06} & \best{0.60} & \best{2.27} & \best{2.93} \\
\texttt{D5} & VibeVoice-ASR-FT & 0.52 & 0.86 & 3.75 & 5.12 \\
\bottomrule
\end{tabular}%
}
\end{table}

\section{Experimental results}
\label{sec:results}
For readability, language and locale names are abbreviated in the following results.
Specifically, \texttt{en-US}, \texttt{en-AU}, \texttt{en-GB}, \texttt{en-PH}, and \texttt{en-IN} denote American, Australian, British, Filipino, and Indian English, respectively.
Similarly, \texttt{fr-CA}, \texttt{pt-BR}, and \texttt{es-MX} denote Canadian French, Brazilian Portuguese, and Mexican Spanish, respectively. 
The remaining labels, \texttt{fr}, \texttt{de}, \texttt{it}, \texttt{ja}, \texttt{ko}, \texttt{pt}, \texttt{ru}, \texttt{es}, \texttt{tl}, \texttt{th}, \texttt{tr}, \texttt{ur}, and \texttt{vi}, denote French, German, Italian, Japanese, Korean, Portuguese, Russian, Spanish, Tagalog, Thai, Turkish, Urdu, and Vietnamese, respectively.

\noindent\textbf{Diarization results:}
Table~\ref{tab:res_der} shows the diarization error rate (DER) results on the development set.
DiariZen-FT (\texttt{D3}) substantially reduced the average DER, clearly outperforming both Baseline$^{\mathrm{1st}}$ (\texttt{D1}) and the original DiariZen model (\texttt{D2}).
Using the VoxBlink2-pretrained SimAMResNet100 speaker embedding model (\texttt{D4}) further improved the average DER, with particularly large gains for some non-English languages such as \texttt{tr}.
Although VibeVoice-ASR-FT (\texttt{D5}) achieved the best DER for several languages/locales, its average DER was higher than that of \texttt{D4}.
This language-dependent behavior may be attributed to its ability to use linguistic and conversational context for speaker attribution, in contrast to the acoustic embedding-based clustering used in the DiariZen systems.

Table~\ref{tab:diar_anal} shows the DER breakdown, which we use to analyze whether the remaining diarization errors are mainly caused by segmentation errors or speaker confusion.
DiariZen-FT (\texttt{D3}) substantially reduced missed speech and false alarms compared with the original DiariZen (\texttt{D2}), indicating that fine-tuning almost eliminated segmentation errors.
The SimAMResNet100-based speaker embedding model (\texttt{D4}) further reduced speaker confusion, but it was still the dominant source of DER.
This suggests that improving speaker attribution is a key direction for future work.
Based on the above results, the cascaded systems evaluated in the following experiments use the diarization outputs from DiariZen-FT with the SimAMResNet100-based speaker embedding model (i.e., \texttt{D4}).

\begin{figure}[t]
  \centering
  \includegraphics[width=0.9\linewidth]{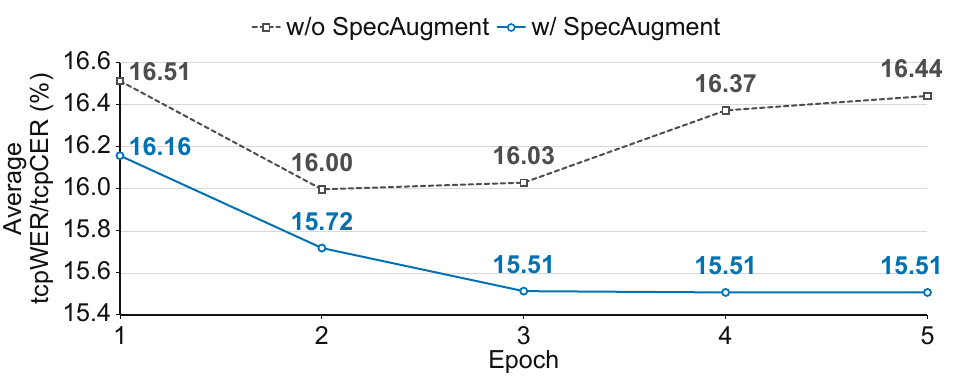}
  \caption{Average tcpWER/tcpCER of Qwen3-ASR-FT across training epochs, with and without SpecAugment.}
  \label{fig:specaug}
\end{figure}

\noindent\textbf{Overall results:}
Figure~\ref{fig:specaug} shows the average tcpWER/tcpCER over training epochs with and without SpecAugment.
Without SpecAugment, the best score was obtained at epoch 2, while SpecAugment consistently improved performance across epochs.
In the following experiments, we therefore use the 2nd-epoch checkpoint for the model without SpecAugment and the 5th-epoch checkpoint for the model with SpecAugment.

Table~\ref{tab:res_tcpwer} summarizes the overall results on the development set.
Compared with the official baseline\urlfootnote{https://github.com/alanshaoTT/MLC-SLM-2nd-Task1-Baseline} (Baseline$^\mathrm{2nd}$, \texttt{S1}), Qwen3-ASR-FT (\texttt{S2}) substantially improved the overall performance.
SpecAugment (\texttt{S3}) further improved the score, and GEC post-processing (\texttt{S4}) achieved the best development-set score of 15.28\%.
Accordingly, we used \texttt{S4} as our final submitted system.
For ablation, we also evaluated a Whisper large-v3 model~\cite{whisper} fine-tuned on the official training data for two epochs with a learning rate of $5\times10^{-6}$ and decoded using beam search with a beam size of 5.
Although this fine-tuned Whisper model (Whisper-FT, \texttt{S6}) outperformed Qwen3-ASR-FT (\texttt{S2}) for some languages, it performed worse on average.

A similar trend was observed on the evaluation set based on the submitted leaderboard results.
The average tcpWER/tcpCER was reduced from 16.70\% to 16.39\% by adding SpecAugment, and was further reduced to 16.27\% by adding GEC.
These results confirm that both SpecAugment and GEC consistently contributed to the final system performance.

On the other hand, the unified VibeVoice-ASR-FT system (\texttt{S5}) also showed promising results.
For several languages/locales, it outperformed the cascaded system, consistent with the diarization result in Table~\ref{tab:res_der}.
This may be because VibeVoice can incorporate linguistic and conversational context into speaker attribution and transcription generation.
However, its average performance was poorer and less stable than that of the cascaded system.
Due to computational constraints, we fine-tuned VibeVoice only on 60-s chunks in this work, and future work should examine whether full-session fine-tuning can improve its robustness and overall performance.

\section{Conclusions}
\label{sec:conclusion}
This paper presented the HINTT submission to the 2nd MLC-SLM Challenge, comparing cascaded and unified approaches for multilingual speaker-attributed ASR.
Our final submission used a cascaded architecture consisting of the fine-tuned DiariZen model, the fine-tuned Qwen3-ASR model, and LLM-based GEC.
We performed all task-specific fine-tuning and model selection using only the official MLC-SLM data, without external data or pseudo-labels.
Experimental results showed that the cascaded system was more reliable than the unified system under our computationally constrained fine-tuning setup.
These findings suggest that modular diarization--ASR pipelines remain strong practical solutions under the current challenge conditions, especially when unified models cannot be fine-tuned on full-session audio.
Unified speech LLMs nevertheless remain promising: with reliable long-form speaker-attributed labels and sufficient resources for full-session fine-tuning, they may enable direct optimization of speaker attribution and ASR over longer conversational contexts.

\newpage
\section{Generative AI Use Disclosure}
A large language model was used only for limited language editing (grammar and readability).
It was not used to generate scientific content or to produce a substantial part of the manuscript.
All authors are responsible and accountable for the paper and consent to its submission.

\bibliographystyle{IEEEtran}
\bibliography{mybib}

@article{qwen3_asr,
  title={Qwen3-{ASR} Technical Report},
  author={Shi, Xian and Wang, Xiong and Guo, Zhifang and Wang, Yongqi and Zhang, Pei and Zhang, Xinyu and Guo, Zishan and Hao, Hongkun and Xi, Yu and Yang, Baosong and others},
  journal={arXiv:2601.21337},
  year={2026}
}

@article{vibevoice_asr,
  title={{VIBEVOICE-ASR} Technical Report},
  author={Peng, Zhiliang and Yu, Jianwei and Chang, Yaoyao and Wang, Zilong and Dong, Li and Hao, Yingbo and Tu, Yujie and Yang, Chenyu and Wang, Wenhui and Xu, Songchen and others},
  journal={arXiv:2601.18184},
  year={2026}
}

@inproceedings{but_mlcslm2025,
  title     = {{BUT} System for the {MLC-SLM} Challenge},
  author    = {Alexander Polok and Jiangyu Han and Dominik Klement and Samuele Cornell and Jan Černocký and Lukáš Burget},
  year      = {2025},
  booktitle = {{Workshop on Multilingual Conversational Speech Language Model (MLC-SLM)}},
  pages     = {23--27},
}

@INPROCEEDINGS{mlcslm2025,
  author={Mu, Bingshen and Guo, Pengcheng and Sun, Zhaokai and Wang, Shuai and Liu, Hexin and Shao, Mingchen and Xie, Lei and Chng, Eng Siong and Xiao, Longshuai and Feng, Qiangze and Wang, Daliang},
  booktitle={ICASSP}, 
  title={Summary on The {Multilingual Conversational Speech Language Model} Challenge: {D}atasets, Tasks, Baselines, and Methods}, 
  year={2026},
  pages={19442--19446},
}

@inproceedings{1st_ranked_mlcslm2025,
  title     = {{ILT}: Iterative {LoRA} Training through Focus–Feedback–Fix for Multilingual Speech Recognition},
  author    = {Qingliang Meng and Hao Wu and Wei Liang and Wei Xu and Qing Zhao},
  year      = {2025},
  booktitle = {{Workshop on Multilingual Conversational Speech Language Model (MLC-SLM)}},
  pages     = {18--22},
}

@inproceedings{unified_mlcslm2025,
  title     = {A Unified Speech {LLM} for Diarization and Speech Recognition in Multilingual Conversations},
  author    = {Phurich Saengthong and Boonnithi Jiaramaneepinit and Sheng Li and Manabu Okumura and Takahiro Shinozaki},
  year      = {2025},
  booktitle = {{Workshop on Multilingual Conversational Speech Language Model (MLC-SLM)}},
  pages     = {56--60},
}

@INPROCEEDINGS{diarizen,
  author={Han, Jiangyu and Landini, Federico and Rohdin, Johan and Silnova, Anna and Diez, Mireia and Burget, Lukáš},
  booktitle={ICASSP}, 
  title={Leveraging Self-Supervised Learning for Speaker Diarization}, 
  year={2025},
  pages={1--5},
}

@INPROCEEDINGS{eend_chunk_filter,
  author={Pálka, Petr and Han, Jiangyu and Delcroix, Marc and Tawara, Naohiro and Burget, Lukáš},
  booktitle={ICASSP}, 
  title={{VBx} for End-to-End Neural and Clustering-Based Diarization}, 
  year={2026},
  pages={17472--17476},
}

@inproceedings{eend_chunk_effect1,
  title     = {Advances in Integration of End-to-End Neural and Clustering-Based Diarization for Real Conversational Speech},
  author    = {Keisuke Kinoshita and Marc Delcroix and Naohiro Tawara},
  year      = {2021},
  booktitle = {Interspeech},
  pages     = {3565--3569},
}

@INPROCEEDINGS{eend_chunk_effect2,
  author={Tawara, Naohiro and Delcroix, Marc and Ando, Atsushi and Ogawa, Atsunori},
  booktitle={ICASSP}, 
  title={{NTT} Speaker Diarization System for {CHiME}-7: {M}ulti-Domain, Multi-Microphone end-to-end and Vector Clustering Diarization}, 
  year={2024},
  pages={11281--11285},
}

@article{eend_chunk_effect3,
  title={Dissecting the segmentation model of end-to-end diarization with vector clustering},
  author={Plaquet, Alexis and Tawara, Naohiro and Delcroix, Marc and Horiguchi, Shota and Ando, Atsushi and Araki, Shoko and Bredin, Herv{\'e}},
  journal={arXiv:2506.11605},
  year={2025}
}

@inproceedings{specaugment,
  title     = {{SpecAugment}: A Simple Data Augmentation Method for Automatic Speech Recognition},
  author    = {Daniel S. Park and William Chan and Yu Zhang and Chung-Cheng Chiu and Barret Zoph and Ekin D. Cubuk and Quoc V. Le},
  year      = {2019},
  booktitle = {Interspeech},
  pages     = {2613--2617},
}

@InProceedings{whisper,
  title = 	 {Robust Speech Recognition via Large-Scale Weak Supervision},
  author =       {Radford, Alec and Kim, Jong Wook and Xu, Tao and Brockman, Greg and McLeavey, Christine and Sutskever, Ilya},
  booktitle = 	 {ICML},
  pages = 	 {28492--28518},
  year = 	 {2023},
  volume = 	 {202},
}

@inproceedings{voxblink2,
  title     = {{VoxBlink2}: {A 100K+} Speaker Recognition Corpus and the Open-Set Speaker-Identification Benchmark},
  author    = {Yuke Lin and Ming Cheng and Fulin Zhang and Yingying Gao and Shilei Zhang and Ming Li},
  year      = {2024},
  booktitle = {Interspeech},
  pages     = {4263--4267},
}

@inproceedings{chime6,
  title     = {{CHiME-6} Challenge: {T}ackling Multispeaker Speech Recognition for Unsegmented Recordings},
  author    = {Shinji Watanabe and Michael Mandel and Jon Barker and Emmanuel Vincent and Ashish Arora and Xuankai Chang and Sanjeev Khudanpur and Vimal Manohar and Daniel Povey and Desh Raj and David Snyder and Aswin Shanmugam Subramanian and Jan Trmal and Bar Ben Yair and Christoph Boeddeker and Zhaoheng Ni and Yusuke Fujita and Shota Horiguchi and Naoyuki Kanda and Takuya Yoshioka and Neville Ryant},
  year      = {2020},
  booktitle = {{6th International Workshop on Speech Processing in Everyday Environments (CHiME 2020)}},
  pages     = {1--7},
}

@inproceedings{chime7,
  title     = {The {CHiME-7 DASR} Challenge: {D}istant Meeting Transcription with Multiple Devices in Diverse Scenarios},
  author    = {Samuele Cornell and Matthew S. Wiesner and Shinji Watanabe and Desh Raj and Xuankai Chang and Paola Garcia and Yoshiki Masuyama and Zhong-Qiu Wang and Stefano Squartini and Sanjeev Khudanpur},
  year      = {2023},
  booktitle = {{7th International Workshop on Speech Processing in Everyday Environments (CHiME 2023)}},
  pages     = {1--6},
}

@inproceedings{chime8,
  title     = {The {CHiME-8 DASR} Challenge for Generalizable and Array Agnostic Distant Automatic Speech Recognition and Diarization},
  author    = {Samuele Cornell and Tae Jin Park and He Huang and Christoph Boeddeker and Xuankai Chang and Matthew Maciejewski and Matthew S Wiesner and Paola Garcia and Shinji Watanabe},
  year      = {2024},
  booktitle = {{8th International Workshop on Speech Processing in Everyday Environments (CHiME 2024)}},
  pages     = {1--6},
}

@inproceedings{speakerlm,
author = {Yin, Han and Chen, Yafeng and Deng, Chong and Cheng, Luyao and Wang, Hui and Tan, Chao-Hong and Chen, Qian and Wang, Wen and Li, Xiangang},
title = {{SpeakerLM}: {E}nd-to-end versatile speaker diarization and recognition with multimodal large language models},
year = {2026},
booktitle = {AAAI},
}

@inproceedings{tagspeech,
    title = "{T}ag{S}peech: {E}nd-to-End Multi-Speaker {ASR} and Diarization with Fine-Grained Temporal Grounding",
    author = "Huo, Mingyue  and
      Shao, Yiwen  and
      Zhang, Yuheng",
    booktitle = "ACL",
    year = "2026",
    pages = "41847--41862",
}

@INPROCEEDINGS{jedis_llm,
  author={Shi, Mohan and Xiao, Xiong and Fan, Ruchao and Ling, Shaoshi and Li, Jinyu},
  booktitle={ICASSP},
  title={Train Short, Infer Long: {Speech-LLM} Enables Zero-Shot Streamable Joint {ASR} and Diarization on Long Audio}, 
  year={2026},
  pages={17442--17446},
}

@article{speaker_reasoner,
  title={{Speaker-Reasoner}: {S}caling Interaction Turns and Reasoning Patterns for Timestamped Speaker-Attributed {ASR}},
  author={Lin, Zhennan and Wang, Shuai and Sun, Zhaokai and Xie, Pengyuan and Xie, Chuan and Liu, Jie and Zhang, Qiang and Xie, Lei},
  journal={arXiv:2604.03074},
  year={2026}
}

@article{chime8_ntt,
title = {Microphone array geometry-independent multi-talker distant {ASR: NTT} system for {DASR} task of the {CHiME-8} challenge},
journal = {Computer Speech \& Language},
volume = {95},
pages = {101820},
year = {2026},
author = {Naoyuki Kamo and Naohiro Tawara and Atsushi Ando and Takatomo Kano and Hiroshi Sato and Rintaro Ikeshita and Takafumi Moriya and Shota Horiguchi and Kohei Matsuura and Atsunori Ogawa and Alexis Plaquet and Takanori Ashihara and Tsubasa Ochiai and Masato Mimura and Marc Delcroix and Tomohiro Nakatani and Taichi Asami and Shoko Araki},
}

@article{survey_multi_asr,
title = {Survey of end-to-end multi-speaker automatic speech recognition for monaural audio},
journal = {Computer Speech \& Language},
volume = {99},
pages = {101925},
year = {2026},
author = {Xinlu He and Jacob Whitehill},
}

@INPROCEEDINGS{saasr_compara,
  author={Kanda, Naoyuki and Xiao, Xiong and Wu, Jian and Zhou, Tianyan and Gaur, Yashesh and Wang, Xiaofei and Meng, Zhong and Chen, Zhuo and Yoshioka, Takuya},
  booktitle={ASRU}, 
  title={A Comparative Study of Modular and Joint Approaches for Speaker-Attributed {ASR} on Monaural Long-Form Audio}, 
  year={2021},
  pages={296--303},
}

@inproceedings{hyporadise,
title={{HyPoradise}: {A}n Open Baseline for Generative Speech Recognition with Large Language Models},
author={Chen Chen and Yuchen Hu and Chao-Han Huck Yang and Sabato Marco Siniscalchi and Pin-Yu Chen and EngSiong Chng},
booktitle={NeurIPS Datasets and Benchmarks Track},
year={2023},
}

@inproceedings{nbest_t5,
  title     = {{N-best T5: Robust ASR} Error Correction using Multiple Input Hypotheses and Constrained Decoding Space},
  author    = {Rao Ma and Mark J. F. Gales and Kate M. Knill and Mengjie Qian},
  year      = {2023},
  booktitle = {Interspeech},
  pages     = {3267--3271},
}

@INPROCEEDINGS{rllm_cf,
  author={Fang, Yangui and Chen, Baixu and Peng, Jing and Li, Xu and Xi, Yu and Zhang, Chengwei and Zhong, Guohui},
  booktitle={ASRU}, 
  title={Fewer Hallucinations, More Verification: {A} Three-Stage {LLM}-Based Framework for {ASR} Error Correction}, 
  year={2025},
  pages={1--7},
}

@INPROCEEDINGS{gensec,
  author={Yang, Chao-Han Huck and Park, Taejin and Gong, Yuan and Li, Yuanchao and Chen, Zhehuai and Lin, Yen-Ting and Chen, Chen and Hu, Yuchen and Dhawan, Kunal and Żelasko, Piotr and Zhang, Chao and Chen, Yun-Nung and Tsao, Yu and Balam, Jagadeesh and Ginsburg, Boris and Siniscalchi, Sabato Marco and Chng, Eng Siong and Bell, Peter and Lai, Catherine and Watanabe, Shinji and Stolcke, Andreas},
  booktitle={SLT}, 
  title={Large Language Model Based Generative Error Correction: {A} Challenge and Baselines For Speech Recognition, Speaker Tagging, and Emotion Recognition}, 
  year={2024},
  pages={371--378},
}

@INPROCEEDINGS{gec_llm,
  author={Yang, Chao-Han Huck and Gu, Yile and Liu, Yi-Chieh and Ghosh, Shalini and Bulyko, Ivan and Stolcke, Andreas},
  booktitle={ASRU}, 
  title={Generative Speech Recognition Error Correction With Large Language Models and Task-Activating Prompting}, 
  year={2023},
  pages={1--8},
}

\end{document}